\documentclass[aps,prb,twocolumn,superscriptaddress,showpacs,english]{revtex4-1}%
\usepackage[T1]{fontenc}
\usepackage{babel}
\usepackage{amsmath}
\usepackage{amssymb}
\usepackage{amsfonts}
\usepackage{bm} 
\usepackage{wasysym} 
\usepackage{graphicx}
\usepackage{multirow} 
\usepackage[caption=false]{subfig}
\usepackage[dvipsnames]{xcolor}
\usepackage{breqn}
\usepackage{mhchem} 
\usepackage[linktocpage=true,   colorlinks=true,    pdfborder={0 0 0},
linkcolor=blue,   citecolor=red,   filecolor=yellow,   urlcolor=blue,
bookmarks,   pdfauthor={}, ]{hyperref}%
\providecommand{\U}[1]{\protect\rule{.1in}{.1in}}
\newcommand{\Graz} {Institute of Theoretical and Computational Physics, Graz
University of Technology, NAWI Graz, 8010 Graz, Austria}

\newcommand{\Washington}{Code 6393, Naval Research Laboratory, Washington, 
DC 20375, USA}
\newcommand{\Rome}{Department of Physics, Sapienza Universit\`a di Roma,
00185 Rome, Italy}
\newcommand{\RomeInst}{Istituto dei Sistemi Complessi (ISC)-CNR, 00185 Rome, 
Italy}
\newcommand{\iro}{IrO$_2$}
\begin{document}
\title{\textit{Ab initio} prediction of a two-dimensional variant of the iridate
IrO$_{2}$}
\author{Andriy Smolyanyuk}
\email{andriysmolyanyuk@gmail.com}
\affiliation{\Graz}
\author{Markus Aichhorn}
\affiliation{\Graz}
\author{I.I. Mazin}
\affiliation{\Washington}
\author{Lilia Boeri}
\affiliation{\Rome}
\affiliation{\RomeInst}
\date{\today}

\begin{abstract}
We propose an insulating two-dimensional phase of IrO$_{2}$, predicted by 
\textit{ab initio} evolutionary algorithms. 
The predicted phase is a van der Waals crystal, in which Ir forms a triangular 
lattice, and is energetically competitive with the metastable spinel phase, 
observed experimentally.
Electronic structure calculations show that the magnetic properties of this
phase are highly nontrivial, with an almost perfect degeneracy of 120$^{\circ
}$ \textit{N\'eel} and $Y$-stripe orders, and unusually soft magnetic moments.
The resulting behavior, which we term \emph{easy plane anisotropy}, is
entirely different from what is realized in previously explored Kitaev honeycomb
lattices. Our results thus suggest that IrO$_{2}$ may be an ideal candidate to
realize highly unusual magnetic properties.

\end{abstract}


\maketitle


\section{Introduction}

Two-dimensional materials represent an ideal platform to investigate exotic
physical phenomena, such as charge-density wave, superconductivity,
topological order, etc., and have a wide range of applicability in different
fields, ranging from coating to two-dimensional (2D) electronics
\cite{Novoselov_2d_2016, Wang_graphene_2017,Fiori_electronics_2014}.
The best known systems are usually
obtained by mechanical exfoliation of bulk van der Waals crystals, such as
graphene \cite{Novoselov_2004} (graphite), $h$-BN \cite{Nagashima_1995},
transition metal dicalchogenides (TMDCs) \cite{Bosi_2015}, or
MXenes \cite{Khazaei_2017}.

Besides investigating the properties of materials derived from known bulk
phases, an emerging trend in the field of 2D materials is to use computational
methods in order to discover, predict and characterize completely new
structures. For example, Ref.~\onlinecite{Ataca_2012} investigated the
stability of single-layer \textit{MX}$_{2}$ transition-metal oxides and
dichalcogenides in honeycomb like structures, data mining of structures listed 
in various databases was employed to filter out possible 2D
materials \cite{Lebegue_2013}, high-throughput computations were used
to determine possible exfoliation of 
experimentally known compounds \cite{Mounet_2018}, and evolutionary algorithms
were used to predict new 2D materials \cite{Oganov_ionic_2009,
Zhou_semimetallic_2014}.

Most 2D materials known up to now are non-magnetic metals, semiconductors or
insulators. The Mermin-Wagner theorem postulates that magnetic order is
forbidden in the two-dimensional Heisenberg model at finite
temperature \cite{Mermin_Wagner_1966}. However, recently magnetism in 2D van
der Waals crystals was discovered \cite{Huang_2017, Gong_2017}, showing that
long-range magnetic order is indeed possible in 2D systems, since magnetic
anisotropy removes the restriction coming from the Mermin-Wagner theorem. This
opens the road to the discovery of many other 2D magnetic systems.

In this work, using evolutionary crystal structure prediction, we identify a
hypothetical 2D phase of IrO$_{2}$, which is a purely 2D system with strongly
anisotropic magnetic interactions, including, but not limited to, the Kitaev 
interaction. 
The Kitaev model~\cite{KITAEV2006} has been studied theoretically by several
authors \cite{Kimchi_kitaev-heisenberg_2014,Chaloupka_hidden_2015,
Chaloupka_magnetic_2016}.
Experimental realizations have been proposed in
several systems, such as, for instance, Na$_{2}$IrO$_{3}$, $\alpha$-Li$_{2}%
$IrO$_{3}$, Li$_{2}$RhO$_{3}$, and $\alpha$-RuCl$_{3}$ \cite{models_kitaev}.
The initial proposal by Jackeli \emph{et al.}~\cite{Jackeli_Khaliullin} was to
realize the Kitaev model on the honeycomb lattice, also noting the possibility 
to apply the model to the triangular lattice formed by edge-shared IrO$_{6}$
octahedra, typical, for instance, in layered \textit{ABO}$_{2}$ compounds where
$A$ and $B$ are alkali and transition metal ions, respectively. 
In all the above proposals, the transition metal sublattice is effectively 
embedded in a three-dimensional framework, while in 2D IrO$_{2}$ the Kitaev 
physics may be realized even on a single monolayer.

Once realized on the triangular lattice, the Kitaev model and more complicated variants,
such as the Heisenberg-Kitaev model with possible extensions by additional 
symmetry-allowed anisotropies, show rich phase diagrams. Apart from 
conventional ordered phases, they include exotic phases with nematic order, 
$\mathbb{Z}_2$-vortex crystal phases 
\cite{becker_spin-orbit_2015,catuneanu_magnetic_2015,rousochatzakis_kitaev_2016},
and the quantum spin-liquid state \cite{li_rare-earth_2015,li_anisotropic_2016}.
Another interesting finding is that quantum order-by-disorder effects are present
in the Kitaev model on the triangular lattice that lead to a selection of the easy 
axes, which in turn reduces the degeneracy of the ground state
\cite{jackeli_quantum_2015}.

Using first-principles calculations based on density functional theory (DFT),
we find that our new 2D-IrO$_{2}$ structure is dynamically stable, van der
Waals bound (i.e., easy to exfoliate), and most intriguingly,
exhibits a highly unusual discrete magnetic frustration, namely, a
120${^{\circ}}$ noncollinear structure with a particular spin orientation with
respect to crystallographic axes is essentially degenerate with a stripe
order, again with a specific spin orientation. This degeneracy cannot be
reproduced by a short-range bilinear coupling, whether isotropic or
anisotropic, and may have ramifications far beyond the scope of this paper.

Even more unusual is the softness of the Ir magnetic moment, which can only
have a full magnetization consistent with $j_{\text{eff}}=1/2$ when the
moments lie in the plane. If rotated away from the plane, the moment rapidly
collapses to essentially zero. This is in some sense similar to the popular $XY$
model, but this similarity is misleading: in the $XY$ model magnetic moments are
strictly restricted within the plane, while in our case they can be rotated
away, but their amplitude rapidly decreases with the angle; such excitations,
impossible in the $XY$ model, are allowed here, so that the system would have
different spin dynamics, different response to magnetic field, and even
different thermodynamics.
Such a system is qualitatively different from usually studied 2D materials and
represents an intriguing and unusual class of magnetic models.

Our paper is structured as follows: In Sec.~\ref{sect:structure} we
introduce the \textit{1T} IrO$_{2}$\ structure, and study its energetics and 
structural stability; in Sec.~\ref{sect:electrons} we describe the effect of crystal
field, spin-orbit coupling and local electronic correlations on the electronic
structure; in the next section, Sec.~\ref{sect:magnetism}, we discuss the unusual
magnetic properties predicted by DFT, and the relevance of different model
Hamiltonians to our results. Finally, in the Appendix we describe the details
of the evolutionary search that led us to discover 2D IrO$_{2}$, and provide
the computational details used for all our calculations.

\section{Crystal Structure and Structural Stability}

\label{sect:structure}

\begin{table}[ptb]
\caption{2D-\iro\ crystal structure:
space group no.~136 ($P\bar{3}2/m1$); the lattice parameters are  
$a$=$b$=$3.16$~\AA, $\alpha$=$\beta$=$90^{\circ}$, $\gamma$=$120^{\circ}$ 
and a sufficiently large vacuum layer was inserted.
}%
\label{table:crystal_structure}%
\centering
\begin{ruledtabular}
\begin{tabular}{lcccc}
  Atom & Site & x & y & z \\
\hline
  Ir & 1a & 0 & 0 & 0 \\
  O  & 2d & $\frac{1}{3}$ & $\frac{2}{3}$ & 0.073 \\
\end{tabular}
\end{ruledtabular}
\end{table}

\begin{figure}[b]
\centering
\begin{subfloat}[Top view \label{fig:1T-top}]{
\includegraphics[width=0.7\linewidth]{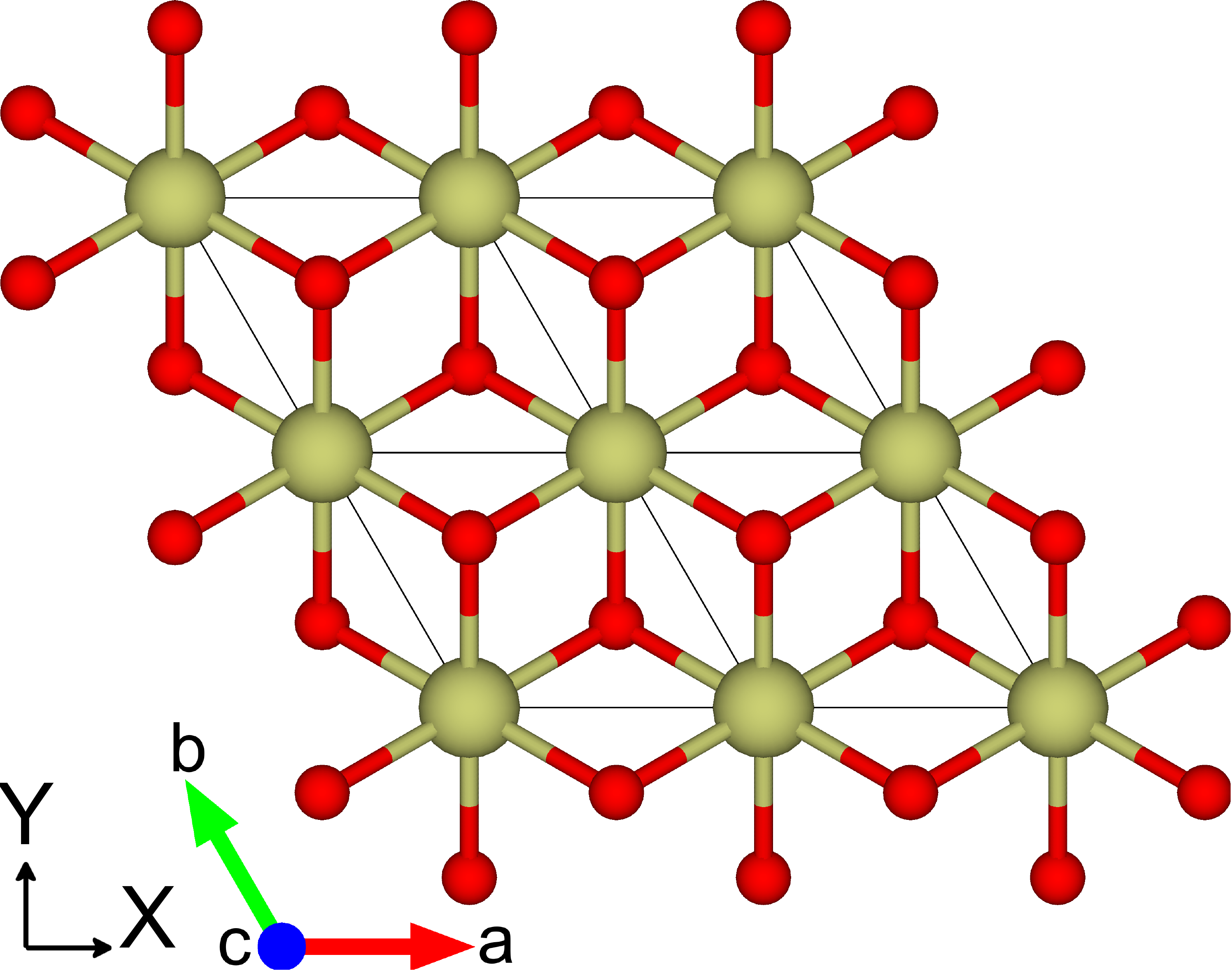}}
\end{subfloat}
\par
\begin{subfloat}[Side view \label{fig:1T-side}]{
\includegraphics[width=0.7\linewidth]{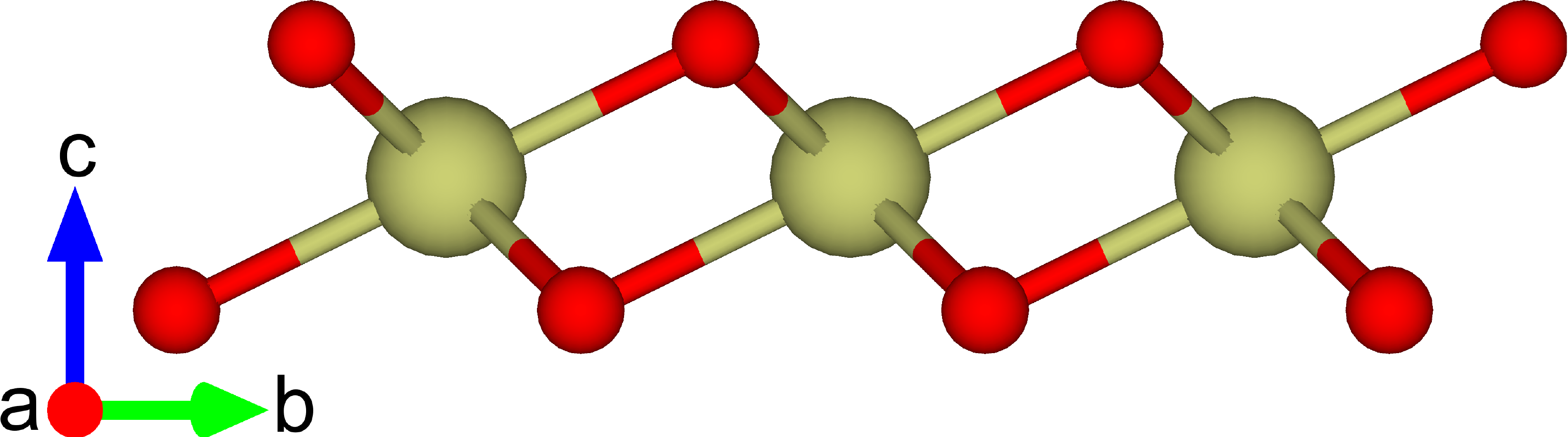}}
\end{subfloat}
  \caption{\textit{1T} IrO$_{2}$\ polytype crystal structure: 
  dark-gray (red) are oxygen atoms, light-gray (green) are iridium atoms; 
  (a) top view, (b) side view. 
  The $X$,$Y$ axes indicate the global ($XYZ$) coordinate system.}%
\label{fig:Structure}%
\end{figure}

The crystal structure of 2D IrO$_{2}$\ is shown in Fig.~\ref{fig:Structure};
it is analogous to the \textit{1T}-polytype of many TMDCs.
In this structure, iridium is
arranged on a triangular lattice, at the centers of O$_{6}$ octahedra, with
slight trigonal distortion (negative) (see the structure parameters in
Table~\ref{table:crystal_structure}).

Our \textit{1T} IrO$_{2}$\ structure, as detailed in the Appendix, was identified
almost accidentally through a sequence of several evolutionary crystal
structure runs, and appears to be a very stable local minimum of the energy
landscape of IrO$_{2}$, where the dominant minima are the two known bulk
phases, with rutile~\cite{rutile} and spinel~\cite{spinel} structures. To the
best of our knowledge, this is the first time that a 2D structure is predicted
for IrO$_{2}$; however, 2D structures that are markedly different from the
bulk structures have been predicted for other binary oxides, such as SiO$_{2}$
and TiO$_{2}$ \cite{SiO2,TiO2}.

Table \ref{table:energies} lists the energetics of different IrO$_{2}%
$\ phases. The stabilization energy of \textit{1T} IrO$_{2}$\ ($\sim$240 meV/atom)
compared to the bulk rutile phase, is in the same ballpark as what is found for
many compounds that exhibit competing layered and bulk phases, such as diamond
and graphite \cite{Diamond_Graphite}, $h$- and cubic BN \cite{BN_and_C}, etc.
Indeed, 1T IrO$_{2}$ appears to be not only more stable energetically than
other 2D polytypes,\cite{polytypes} but also compared to surface terminations
of the rutile ground-state structure and to the first metastable bulk phase,
i.e. spinel; thus, it could be very likely synthesized experimentally.

\begin{table}[ptb]
\caption{Comparison of energies of various 2D IrO$_{2}$\ phases (nonmagnetic
calculations including spin-orbit coupling). }%
\label{table:energies}
\begin{ruledtabular}
\begin{tabular}{lc}
\textbf{Name}           & \textbf{Energy/Atom (meV)}   \\
\hline
Rutile (bulk)           & 0                            \\ \hline
1T                      & 237                          \\
Spinel (bulk)           & 266                          \\
Rutile (001)            & 326                          \\
Rutile (110)            & 385                          \\
1H                      & 621                          \\
\end{tabular}
\end{ruledtabular}
\end{table}

According to our calculation, \textit{1T} IrO$_{2}$ is stable in both monolayer and
bulk form; in fact, our calculations suggest that the latter becomes the
ground state at $\sim$160\,GPa, overcoming the two 3D bulk structures. This
bulk layered phase is a van der Waals crystal, as can be demonstrated by
optimizing with and one without a van der Waals correction (we used the one
from Ref.~\onlinecite{Tkatchenko-Scheffler}, but it is well-known that
alternative choices yield similar results). We found that the interlayer
distance in the two cases differs by 3.3\,\AA \ (9.65 vs. 6.35\,\AA ), which
is larger than for representative transition metal dichalcogenides ($7.4-6.0=
1.4$\,\AA \ for MoS$_{2}$, $6.9-6.0=0.9$\,\AA \ for NbSe$_{2}$), and similar to
that in graphite ($8.8-6.6=2.2$\,\AA) \cite{Bucko_2013}. This means that, once
synthesized, \textit{1T} IrO$_{2}$ will be easily exfoliable.

After the structure was determined, we checked its dynamical stability by
calculating the phonon dispersion. We used a finite displacement method as
implemented in the \texttt{Phonopy} package \cite{phonopy}, using a 2x2x1
supercell. Under realistic conditions, i.e., including spin-orbit coupling 
(SOC), strong correlations and magnetism (see next paragraph), \textit{1T} 
IrO$_{2}$  is dynamically stable.

\section{Electronic Structure}

\label{sect:electrons} Ir$^{4+}$ has one unpaired electron
so one expects the ground state to be magnetic. As
commonly done for Ir$^{4+}$ compounds, we have included electronic correlations using
the GGA+$U$ method with $U-J=2$\,eV, a typical value for such
systems \cite{Winter2016,Liu2015, Li2015}.

\begin{figure}[ptb]
\centering
\includegraphics[width=\linewidth]{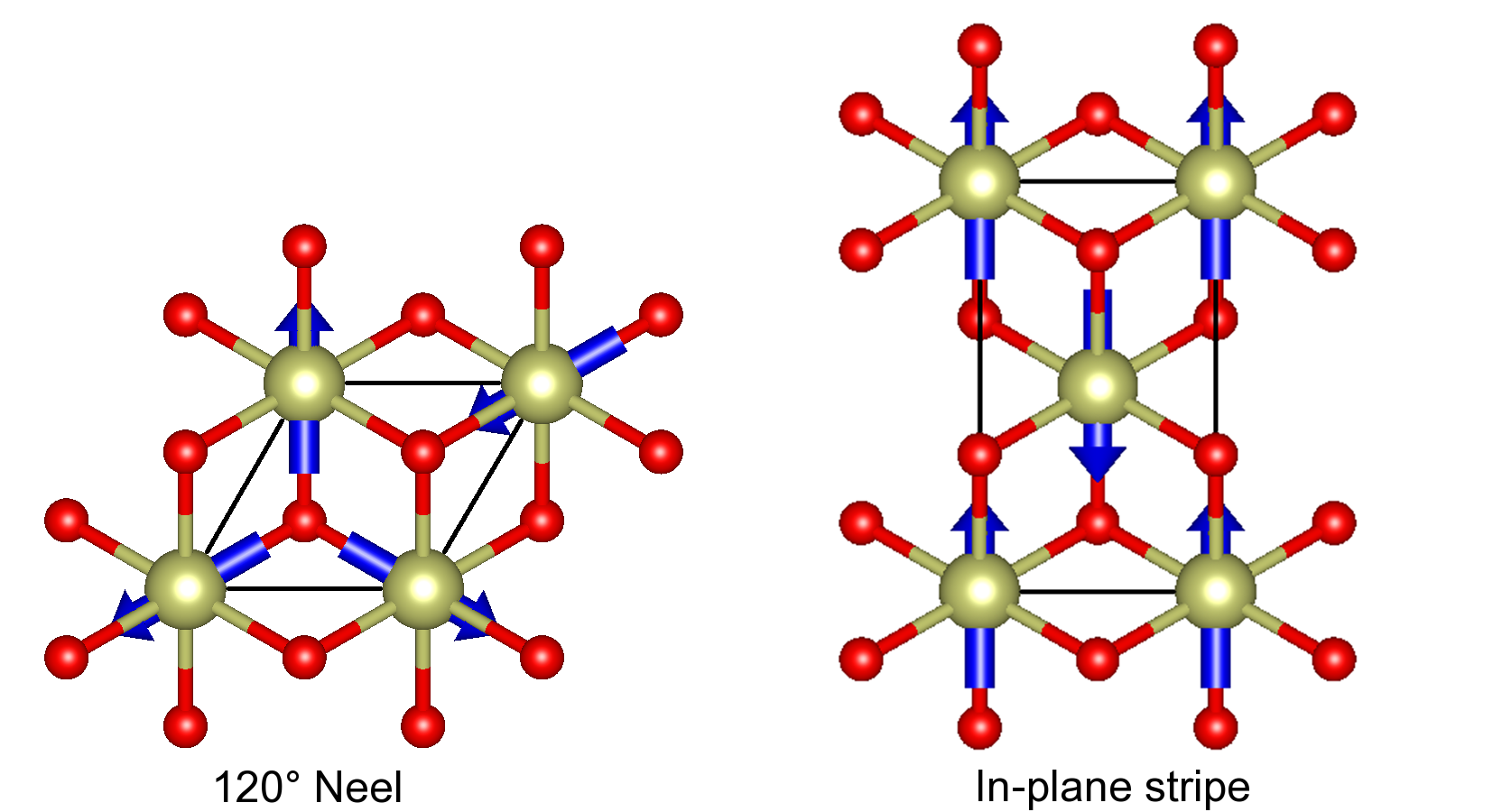} \caption{The
  two nearly degenerate \textit{1T} IrO$_{2}$ magnetic configurations, 
  which are lowest in energy. 
  The arrows show the directions of the spin magnetic moments on the
Ir atoms in the $XY$ plane.}%
\label{fig:Magnetic_configurations}%
\end{figure}

We find two completely different magnetic ground states, degenerate on the
level of computational accuracy of $\lesssim1$ meV/Ir. These two
configurations are the 120$^{\circ}$-N\'eel orientation and an \textit{in-plane
stripe} order with magnetic moments directed along the global $Y$ axis, which we
will denote as \textit{Y stripes}, as shown in
Fig.~\ref{fig:Magnetic_configurations}. The 120$^{\circ}$-Neel configuration
is formed by three sublattices oriented at a 120$^{\circ}$ angle with respect
to each other, and the \textit{Y} stripes configuration is formed by
antiferromagnetically coupled rows of collinear spins lying in the layer's plane.
Our calculations show no qualitative changes in the properties of 
\textit{1T} IrO$_{2}$\ for $1.3<U-J<4.0$\,eV, and the abovementioned degeneracy 
of magnetic phases is observed in this region.

In both cases the system forms a weak Mott-Hubbard insulator; using
$U-J=2.0$\,eV, the band gap for
the \textit{120$^{\circ}$}-N\'eel structure is 0.42\,eV and for the
\textit{Y} stripes it is 0.24\,eV.

\begin{figure}[ptb]
  \centering
\includegraphics[width=\linewidth]{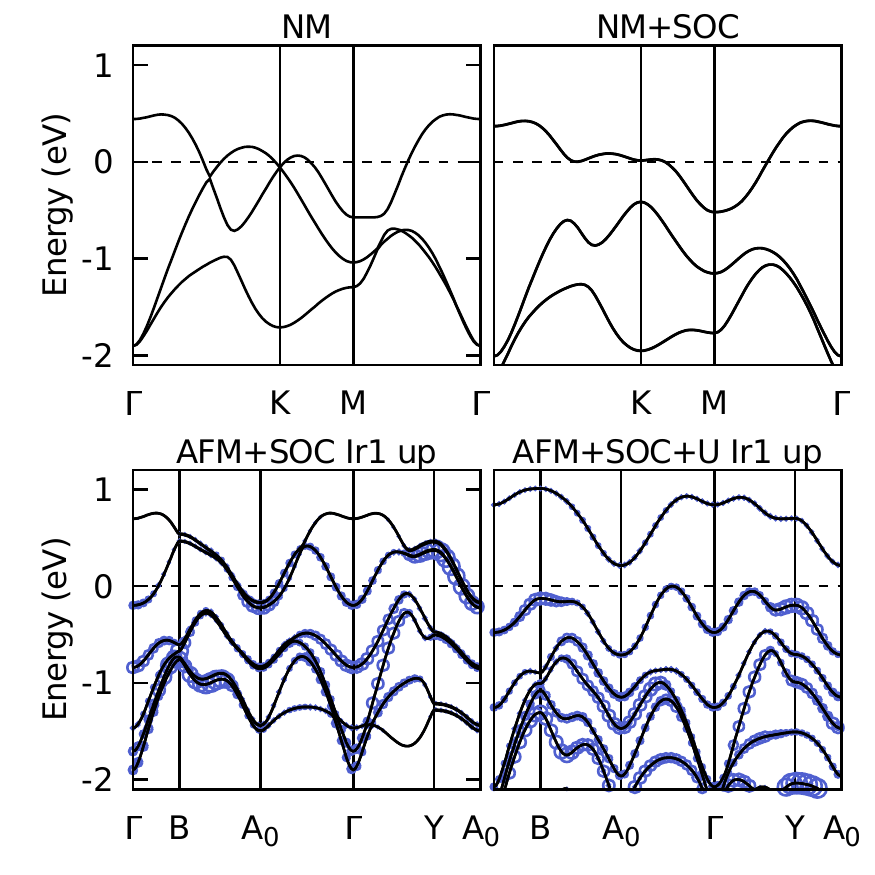}
  \caption{Band structure of \textit{1T} IrO$_{2}$\ states for nonmagnetic IrO$_{2}%
$\ with and without spin-orbit coupling (top panel), and the \textit{Y}-stripe AFM
configuration with SOC and with or without $U=2.7$\,eV (bottom panel).
The bottom panels show the projection on the single Ir spin-up states 
($m_s=\frac{1}{2}$ states) with quantization axis in the $Y$-direction. 
Calculations are done with \texttt{Wien2k} \cite{wien2kfootnote}.
The high-symmetry $k$-points are: (upper panel) $\Gamma(0,0,0)$,
$K(\frac{1}{3},\frac{1}{3},0)$, $M(\frac{1}{2},0,0)$ for trigonal cell
($a$=$b$=3.16\,\AA, $\alpha$=$\beta$=90$^\circ$, $\gamma$=120$^\circ$)
and (bottom panel) $\Gamma(0,0,0)$, $B=(\frac{1}{2},0,0)$, $Y=(0,\frac{1}{2},0)$,
$A_{0}=(\frac{1}{2},\frac{1}{2},0)$ for monoclinic supercell ($a$=3.16\,\AA,
$b$=5.47\,\AA, $\alpha=\beta=\gamma$=90$^\circ$).}%
\label{fig:Fatbands}%
\end{figure}

The origin of the gap is the same as in other Ir$^{4+}$ iridates. The $t_{2g}$
splits into a doublet and a quartet, reminiscent of the $j_{\text{eff}}=1/2$ and 
$j_{\text{eff}}=3/2$ spin-orbit driven splitting of a $t_{2g}$ level..
The doublet forms a narrow half-filled band that can be easily split by a
moderate Hubbard interaction \cite{Kim_2008}. As one can see in the
lower panel of 
Fig.~\ref{fig:Fatbands}, already for $U=0$ the exchange splitting is trying to open
the gap, but it is too weak. Adding a finite $U$ eventually opens up a
gap.

\section{Magnetic Structure}

\label{sect:magnetism} 
\begin{figure}[ptb]
\centering
\includegraphics[width=\linewidth]{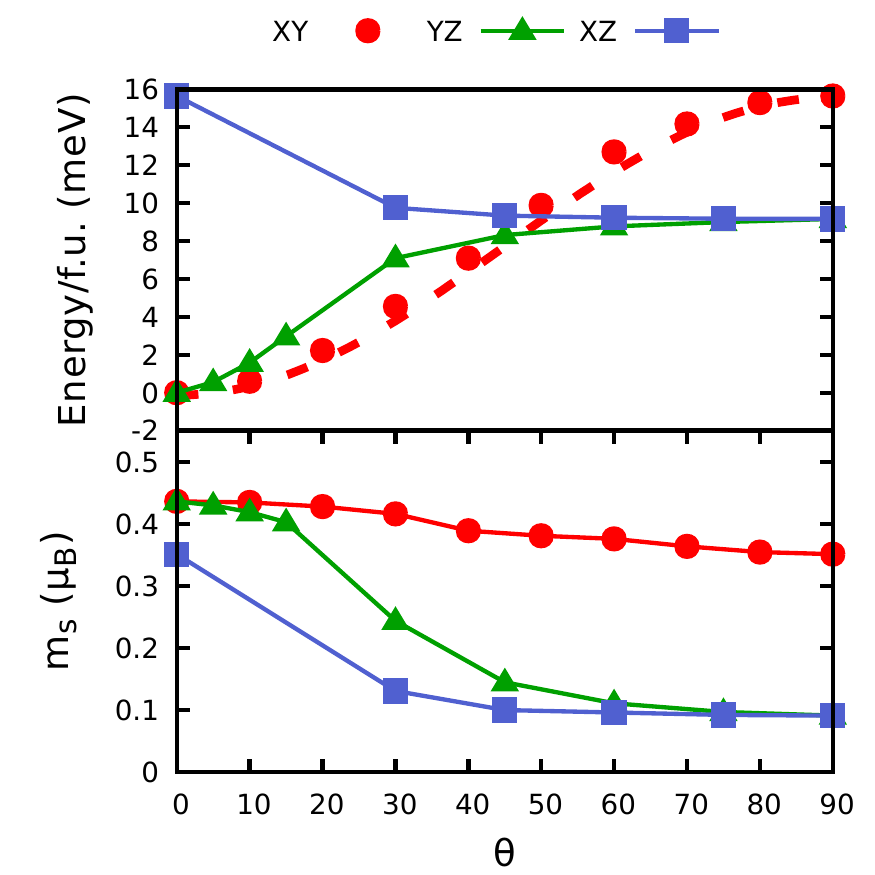}\caption{ Dependence
of the energy (top panel) and the spin magnetic moment on the Ir site (bottom
panel) on the rotation angle for stripe order, with spins rotating in the $XY$
(circles), $YZ$ (triangles) and $XZ$ (squares) planes. The dashed line is the
fit of energy of $XY$ rotations to the model defined by Eq.~\eqref{eq:H2D}.
For rotations in the $XY$ plane, $\theta$ is the in-plane rotation angle, and
$\theta=0$ corresponds to the $Y$ direction; for rotations in the $YZ$ and
$XZ$ planes, the angle $\theta$ is the out-of-plane rotation angle, with
$\theta=0$ indicating the in-plane direction. Solid lines are meant as a guide
to the eye.}%
\label{fig:Stripes}%
\end{figure}

Magnetic interactions are strongly anisotropic (see
Table~\ref{table:magnetic} and top panel of Fig.~\ref{fig:Stripes}). For
rotations of the stripe order out of the $XY$ plane the minimum of the energy
corresponds to the in-plane direction, while for rotations in the $XY$ plane
the $Y$ direction, orthogonal to an Ir-Ir bond, is energetically more
favorable. Importantly, as Fig.~\ref{fig:Stripes} shows, as magnetic moments
are rotated away from the $XY$ plane by more than approximately 15$^{\circ}$, they
collapse from 0.4-0.5\,$\mu_{B}$ to 0.1\,$\mu_{B}$ or less (numbers
are given for spin moments; the
total moments drop from about 1.0 to 0.2-0.25\,$\mu_{B}$).

It has become customary to use the following anisotropic exchange model, often 
referred to it as an extended Heisenberg-Kitaev (EHK) Hamiltonian, for 
describing magnetic interactions in iridates.
It can be conveniently written using the form suggested in 
Refs.~\onlinecite{Maksimov_2019,Chaloupka_hidden_2015, Jackeli_Khaliullin}:
\begin{align}
H= &  \sum_{\langle ij\rangle}J\left(  S_{i}^{x}S_{j}^{x}+S_{i}^{y}S_{j}%
^{y}+\Delta S_{i}^{z}S_{j}^{z}\right)  \label{eq:EHK}\\
&  +2J_{\pm\pm}\left(  (S_{i}^{x}S_{j}^{x}-S_{i}^{y}S_{j}^{y})c_{\alpha
}-(S_{i}^{x}S_{j}^{y}+S_{i}^{y}S_{j}^{x})s_{\alpha}\right)  \nonumber\\
&  +J_{z\pm}\left(  (S_{i}^{y}S_{j}^{z}+S_{i}^{z}S_{j}^{y})c_{\alpha}%
-(S_{i}^{x}S_{j}^{z}+S_{i}^{z}S_{j}^{x})s_{\alpha}\right) \nonumber.
\end{align}
The sum above is over nearest-neighboring sites $i$ and $j$; $c_{\alpha}
=\cos{\varphi_{\alpha}}$ and $s_{\alpha}=\sin{\varphi_{\alpha}}$, where
$\varphi_{\alpha}=\{0,\frac{2\pi}{3},-\frac{2\pi}{3}\}$ is the bond angle
between the direction of the $ij$ bond and the $X$ axis. 

An interesting issue, not always brought to light, is the physical
meaning of the operators $S$ here. 
Given a single (Kramers-degenerate, in absence of a 
magnetic field) band, separated from the rest of the $t_{2g}$ manifold,
it is always possible to assign to it a pseudospin quantum number 
that would have the same property as a spin-1/2 operator. The beauty 
of Eq.~\eqref{eq:EHK}, as applied to materials such as Na$_2$IrO$_3$ or
RuCl$_3$, is that this pseudospin has a direct, simple and physically
transparent interpretation in terms of $j_{eff}=1/2$ states. The latter have the spin 
and orbital moment collinear, adding up to $M=M_s+M_l=1$\,$\mu_B$, so that 
the direction of $S$ coincides with the directions of $M$ and $M_s$, and the $g$-factor
is isotropic and equal to 2. This is, approximately, the case in 
Na$_2$IrO$_3$, but as we will discuss later, not entirely so in our compound.

Exchange parameters in Eq.~\eqref{eq:EHK} can also be rewritten
in the notations of Ref.~\onlinecite{Chaloupka_hidden_2015} as 
$J_{\pm\pm}=\frac {1}{6}(2\Gamma^{\prime}-2\Gamma-K)$,
$J_{z\pm}=\frac{\sqrt{2}}{3} (K-\Gamma+\Gamma^{\prime})$,
$J\Delta=\frac{1}{3}(3\widetilde{J}+K+2\Gamma+4\Gamma ^{\prime})$
(since the same symbol $J$ is used in Refs.~\onlinecite{Chaloupka_hidden_2015} 
and \onlinecite{Maksimov_2019} we substitute it by $\widetilde{J}$ when
we talk about notations from Ref.~\onlinecite{Chaloupka_hidden_2015}).
While the easy-plane anisotropy is not affected by pure Kitaev
interactions, it is affected by the non-Kitaev terms $\Gamma$ and $\Gamma^{\prime}$. 
$J$ is the bond-independent isotropic magnetic interaction term, $J_{\pm\pm}$ 
(anisotropy in the $XY$ plane) and $J_{z\pm}$ (anisotropy in the $YZ$ plane) are 
the bond-dependent anisotropic terms and $\Delta$ is what is
usually called the Ising exchange.
For collinear spins, the contribution of $J_{z\pm}$ cancels out, and
an easy-plane anisotropy is solely determined by $\Delta$;
$\Delta>1$ corresponds to an easy
$Z$ axis, and $\Delta<1$ to an $XY$ easy-plane anisotropy. However,
for other magnetic configurations an easy-plane 
vs. easy-axis anisotropy involves also $J_{z\pm}$, whereas
$J_{\pm\pm}$ produces an anisotropy within the $XY$-plane.

The easy-plane anisotropy that we see in the calculations is, on the first
glance, an indication of a sizable $\Delta$ with $J_{z\pm}\rightarrow 0$,
since nonzero $J_{z\pm}$ favors out-of-plane direction.
However, a closer look reveals that in our case the most important contribution
to the anisotropy does not come from $\Delta,$ but from the near collapse of the 
magnetic moment  for out-of-plane directions (see the bottom panel of 
Fig.~\ref{fig:Stripes}). This observation is in direct contradiction with the
$j_{eff}=1/2$ description, and can only be reconciled with a description in
terms of pseudospins if a highly anisotropic and/or strongly nondiagonal
$g$-tensor is assumed.
Direct inspection of the band structure 
in a collapsed moment state indicates that the split-off hole band changes its 
spin polarization as a function of $k$ and the latter integrates to nearly zero.
This is the effect of strong hybridization of the hole band with other
electronic states that prevents us from describing it as a nearly-pure 
$j_{eff}=1/2$ state. Not having access to the pseudospin direction
(note that in either experiment or DFT calculations only the  
magnetic moments are observed, and the pseudospins, being a mathematical 
construct, are not directly accessible) we cannot therefore
evaluate the parameters of the Hamiltonian Eq.~\eqref{eq:EHK} for
the situations where the actual magnetic moments collapse.

On the other hand, as long as only in-plane spins are considered,
the pseudospin description is not that far from the $j_{eff}=\frac{1}{2}$
description; the calculated spin and orbital moments are indeed parallel to each other
[and this direction can be taken as the direction of $S$ in Eq.~\eqref{eq:EHK}]. 
They add up to numbers consistent with $M=1$\,$\mu_B$ (with a reasonable
hybridization reduction), and thus can be used in Eq.~\eqref{eq:EHK} in a 
meaningful way:
\begin{align}
H= &  \sum_{\langle ij\rangle}J\left(  S_{i}^{x}S_{j}^{x}+S_{i}^{y}S_{j}%
^{y}\right)  \label{eq:H2D}\\
+ &  2J_{\pm\pm}\left(  (S_{i}^{x}S_{j}^{x}-S_{i}^{y}S_{j}^{y})c_{\alpha
}-(S_{i}^{x}S_{j}^{y}+S_{i}^{y}S_{j}^{x})s_{\alpha}\right)  .\nonumber
\end{align}


The classical per-site energies of the phases with magnetization directions in
the $XY$ plane, namely the 120$^{\circ}$ order, rotations of the stripe order
in the $XY$ plane and an in-plane FM order (see Fig.~\ref{fig:Stripes} and
Table~\ref{table:magnetic}) are
\begin{align}
E_{120}= &  E_{0}-\frac{3}{2}J,\\
E_{XY}^{Stripe}= &  E_{0}-J-4J_{\pm\pm}\cos{2\theta},\\
E_{FM}= &  E_{0}+3J,
\end{align}
where $\theta=0$ corresponds to the $Y$-direction.
We want to note that the energy of inplane 120$^{\circ}$ order does not
change upon the rotation around $Z$ axis.
However, out-of-plane variants of 120$^\circ$ order are higher in energy.

The least-squares fit of $J$, $J_{\pm\pm}$ and $E_{0}$ to the available data
is shown in Table~\ref{table:fit}, while the energies predicted by the fitted
model for different magnetic configurations are shown in the last column of
Table~\ref{table:magnetic} ($E_{fit}$). The agreement in this case is quite
good, indicating that the in-plane restricted model is a rather good match for
DFT energies.

The anisotropic $J_{\pm\pm}$ term is only three times smaller than the
isotropic $J$ term, emphasizing strong \textit{in-plane} anisotropy. This is
in addition to the above-mentioned nontrivial anisotropy stemming from the
softness of magnetic moments. The degeneracy of the 120$^{\circ}$ N\'eel and the
$Y$ stripe configurations is only reproduced within $4$ meV, while in the
calculations the two configurations are degenerate within a fraction of meV.
Such an accidental discontinuous degeneracy, a consequence of the intrinsic
magnetic frustration, adds a further exciting aspect to the unusual spin
dynamics of this system.

\begin{table}[ptb]
\caption{Energy and absolute values of the spin and orbital magnetic moments
for different magnetic phases. The direction of the magnetic moment is denoted
in the global coordinate system; the last column shows energies (per formula
unit) estimated using the in-plane Hamiltonian Eq.~\eqref{eq:H2D} with the
parameters from Table~\ref{table:fit}. Note that the energy fits are
not applicable to out-of-plane directions, lines 3 and 7.}%
\label{table:magnetic}%
\centering
\begin{ruledtabular}
\begin{tabular}{lcccc}
Name & Energy/f.u. (meV) &
$m_s$ ($\mu_{B}$) &
$m_l$ ($\mu_{B}$) & E$_{fit}$ (meV) \\
\hline
120 Neel &   0               & 0.40  & 0.50 & 4  \\
Y-stripe &   0               & 0.43  & 0.55 & 0  \\
Z-stripe &   9               & 0.09  & 0.11 & -  \\
X-stripe &  16               & 0.35  & 0.50 & 16 \\
Y-FM     &  34               & 0.47  & 0.61 & 35 \\
X-FM     &  34               & 0.47  & 0.61 & 35 \\
Z-FM     &  49               & 0.09  & 0.16 & -  \\
\end{tabular}
\end{ruledtabular}
\end{table}

\begin{table}[ptb]
\caption{Parameters of the restricted Heisenberg-Kitaev Hamiltonian
Eq.~\eqref{eq:H2D}, obtained by a least-square fit.}%
\label{table:fit}%
\begin{ruledtabular}
\begin{tabular}{ccc}
\text{$J$} (meV) & \text{$J_{\pm\pm}$} (meV) & \text{$E_0$} (meV) \\
\hline
6.7  $\pm$ 0.4 & 2.0 $\pm$ 0.2  &  14.4 $\pm$ 0.5\\
\end{tabular}
\end{ruledtabular}
\end{table}

\section{Conclusions}

We predicted a hypothetical insulating layered phase of IrO$_{2}$\ with
a triangular Ir$^{4+}$ lattice -- \textit{1T} IrO$_{2}$. We show that it is likely to
be a very stable phase, albeit not the most stable phase. The bonding between the
layers is very weak, so monolayers could be easily obtained by exfoliation.
The Ir$^{4+}$ ions are in an (approximately) $j_{\text{eff}}=1/2$ state, as
are most other Ir$^{4+}$ iridates, too. Our \textit{ab initio} calculations reveal
several highly unusual aspects of magnetism in this system: (i) a nontrivial
accidental degeneracy of the magnetic ground state: the \textit{120$^{\circ}$
N\'eel} orientation and the $Y$ stripe (in-plane orientation perpendicular to
the Ir-Ir bonds) are degenerate down to 1 meV/atom; (ii) a strong magnetic
anisotropy even within the hexagonal plane, a reflection of strong
bond-dependent magnetic exchange interactions; (iii) finally, the calculated
moments are exceptionally soft in the sense that they essentially disappear if
forced to tilt away from the plane. This behavior can be called an
\textquotedblleft easy-plane anisotropy,\textquotedblright but this anisotropy
is physically very distinct from the \textquotedblleft
conventional\textquotedblright\ easy-plane behavior, provided by either
single-site or exchange anisotropy. Thus, despite the same geometry as
honeycomb iridates, in \textit{1T} IrO$_{2}$\ Kitaev interaction does not strongly
dominate the magnetic physics. This system is entirely different from
previously explored Kitaev honeycomb lattices, and promises nontrivial and yet
unexplored magnetic properties.



\begin{acknowledgments}
We acknowledge funding from the Austrian Science Fund FWF through SFB ViCoM,
Project No. F04115 and START program Y746, and computational resources from the
VSC3 of the Vienna University of Technology and HPC TU Graz. I.I. Mazin
acknowledges support by ONR through the NRL basic research program and
Sapienza University of Rome through Bando Professori Visitatori 2018.
\end{acknowledgments}

\section*{Appendix: Computational Details}
\subsection*{Details of the crystal structure search}

Our structure was identified through a sequence of several evolutionary
crystal structure runs, performed with the \texttt{USPEX}
package~\cite{uspex,uspex2,uspex3} employing four-step structural relaxation.
First, in an unbiased 3D structure search, we realized that metastable 2D
structures of IrO$_{2}$, with stabilization energies of 200 meV, coexisted
with the known bulk structures. For a further refinement we repeated the
search with different accuracies, but limiting our search to 2D structures.
This yielded \textit{1T} as the most stable polytype \cite{polytypes}, also compared to
(001) and (110) cuts of the rutile surface; see Table~\ref{table:energies}.

\subsection*{Details of DFT calculation}

To calculate the total energies and perform structural optimization we used
density functional theory (DFT) in the generalized gradient approximation
(GGA) with the Perdew-Burke-Ernzerhof functional \cite{GGA1, GGA2} as implemented
in the \texttt{VASP} package \cite{VASP1, VASP2, VASP3, VASP4} using the
projector augmented wave method (PAW).\cite{PAW1, PAW2}

For the post-processing of the results from the evolutionary search and
calculation of properties the energy cutoff was set to 600~eV, and a $\Gamma
$-centered Monkhorst-Pack grid~\cite{MP1, MP2} with the reciprocal-space
resolution of 0.023~$2\pi\mbox{\r{A}}^{-1}$ was used. For the magnetic
calculations the Wigner-Seitz radius was set to 1.423 and 0.9~\AA \ for
Ir and O atoms respectively; the penalty term $\lambda$ for constrained
magnetic calculations was set to the value 10.

The band structure plots shown in Fig.~\ref{fig:Fatbands} were obtained with
\texttt{Wien2k}.\cite{wien2k} Due to different implementations of the
GGA+$U$ method in \texttt{VASP} and \texttt{Wien2k}, the value of $U$ is
not directly transferable between the two codes. For that reason, we
had to use a slightly larger value of $U=2.7$\,eV when using
\texttt{Wien2k}. This value yields a similar gap as for VASP calculations
with $U=2.0$\,eV. For the \texttt{Wien2k} calculations we used a
16x16x1 k-mesh, defined on a $\Gamma$-centered point grid \cite{Blochl_1994}.

\bibliographystyle{apsrev4-1}

%
\end{document}